\documentclass[11pt]{article}
\usepackage{amsfonts}
\usepackage{amssymb}
\usepackage{amsmath}
\usepackage{a4wide}
\usepackage{graphics}

\newtheorem{theorem}{Theorem}

\newtheorem{corollary}{Corollary}

\newtheorem{proposition}{Proposition}

\makeatletter
\def\@removefromreset#1#2{\let\@tempb\@elt
     \def\@tempa#1{@&#1}\expandafter\let\csname @*#1*\endcsname\@tempa
     \def\@elt##1{\expandafter\ifx\csname @*##1*\endcsname\@tempa\else
    \noexpand\@elt{##1}\fi}     \expandafter\edef\csname cl@#2\endcsname{\csname cl@#2\endcsname}     \let\@elt\@tempb
     \expandafter\let\csname @*#1*\endcsname\@undefined}

\@removefromreset{equation}{section}

\@removefromreset{theorem}{section}
\makeatother
\input{tcilatex}
\begin{document}

\title{Upper bounds on violation of Bell-type inequalities by a multipartite
quantum state}
\author{Elena R. Loubenets\bigskip \\
Applied Mathematics Department, Moscow State Institute \\
of Electronics and Mathematics, Moscow 109028, Russia}
\maketitle

\begin{abstract}
We present the new exact upper bounds on the maximal Bell violation for the
generalized $N$-qubit GHZ state, the $N$-qudit GHZ state and, in general,
for an arbitrary $N$-partite quantum state, possibly infinite-dimensional.
Our results indicate that, for an $N$-partite quantum state of any Hilbert
space dimension, violation of any Bell-type inequality (either on
correlation functions or on joint probabilities) with $S$ settings and any
number of outcomes at each site cannot exceed $(2S-1)^{N-1}$.
\end{abstract}

\section{Introduction}

Multipartite Bell-type inequalities\footnote{%
On the general framework for multipartite Bell-type inequalities, see \cite%
{1}.} are now widely used in many schemes of quantum information processing.
However, the exact upper bounds on quantum Bell violations are well known
only in case of bipartite correlation Bell-type inequalities where,
independently on a Hilbert space dimension of a bipartite quantum state and
numbers of measurement settings per site, quantum violations cannot exceed 
\cite{2, 3} the Grothendieck constant.

Bounds on violation by a bipartite quantum state of Bell-type inequalities
for joint probabilities have been recently intensively discussed in the
literature both computationally \cite{4} and theoretically, see \cite{5, 6,
7} and references therein.

For an arbitrary $N$-partite quantum state, the exact upper bounds on the
maximal quantum Bell violation have not been reported in the literature but
it has been argued in \cite{5} that, via increasing of a Hilbert space
dimension of some tripartite quantum states, these states \textquotedblleft
can lead to arbitrarily large violations of Bell
inequalities\textquotedblright \footnote{%
Cited according to \cite{5}}.

In this concise presentation on our results in [8-10], we present the exact
upper bounds on violation by $N$-partite quantum states of any Bell-type
inequality, either on correlation functions or on joint probabilities.
Specified for $N=2,3,$ our new general results improve the bipartite upper
bounds reported in \cite{6, 7} and also clarify the range of applicability
of the tripartite lower estimate found in \cite{5}.

\section{Some new Hilbert space notions}

In this section, we shortly introduce some new tensor Hilbert space notions
[8-10] needed for our further consideration.

\underline{\textbf{Source operators}}\textbf{. }For a state $\rho $ on a
complex separable Hilbert space $\mathcal{H}_{1}\otimes \cdots \otimes 
\mathcal{H}_{N},$ denote by $T_{S_{1}\times \cdots \times S_{N}}^{(\rho )}$
any of its self-adjoint trace class dilations to space $\mathcal{H}%
_{1}^{\otimes S_{1}}\otimes \cdots \otimes \mathcal{H}_{N}^{\otimes S_{N}}$.

We refer to dilation $T_{S_{1}\times \cdots \times S_{N}}^{(\rho )}$ as 
\emph{an }$S_{1}\times \cdots \times S_{N}$\emph{-setting source operator}
for state $\rho $ and set $T_{_{1\times \cdots \times 1}}^{(\rho )}:=\rho .$
For any source operator $T$, it trace $\mathrm{tr}[T]=1.$

\begin{proposition}
For any $N$-partite quantum state $\rho ,$ possibly infinite-dimensional,
and any positive integers $S_{1},...,S_{N}\geq 1,$ an $S_{1}\times \cdots
\times S_{N}$-setting source operator $T_{S_{1}\times \cdots \times
S_{N}}^{(\rho )}$ exists.
\end{proposition}

\medskip

\underline{\textbf{Tensor positivity}}\textbf{.}{\large \ }We refer to a
trace class operator $W$ on a Hilbert space space $\mathcal{G}_{1}\otimes 
\mathcal{\cdots }\otimes \mathcal{G}_{m},$ $m\geq 1$ as \emph{tensor} \emph{%
positive} and denote this by $W\overset{\otimes }{\geq }0$ if 
\begin{equation}
\mathrm{tr}[W\{X_{1}\otimes \cdots \otimes X_{m}\}]\geq 0,  \label{0}
\end{equation}%
for any positive bounded linear operators $X_{1},...,X_{m}$ on spaces $%
\mathcal{G}_{1},..,\mathcal{G}_{m}$, respectively.\bigskip

\underline{\textbf{The covering norm}}\textbf{. }For a self-adjoint trace
class operator $W$ on $\mathcal{G}_{1}\otimes \mathcal{\cdots }\otimes 
\mathcal{G}_{m}$, we call a tensor positive trace class operator $W_{cov}$\
on $\mathcal{G}_{1}\otimes \mathcal{\cdots }\otimes \mathcal{G}_{m}$
satisfying relations%
\begin{equation}
W_{cov}\pm W\overset{\otimes }{\geq }0,  \label{01}
\end{equation}%
as \emph{a trace class covering} \emph{of} $W.$

\begin{proposition}
For any operator\footnote{%
Here, $\mathcal{T}_{\mathcal{G}_{1}\otimes \mathcal{\cdots }\otimes \mathcal{%
G}_{m}}$ and $\mathcal{T}_{\mathcal{G}_{1}\otimes \mathcal{\cdots }\otimes 
\mathcal{G}_{m}}^{(sa)}$ denote, correspondingly, the space of all trace
class operators and the space of all self-adjoint trace class operators on $%
\mathcal{G}_{1}\otimes \mathcal{\cdots }\otimes \mathcal{G}_{m}$.} $W\in 
\mathcal{T}_{\mathcal{G}_{1}\otimes \mathcal{\cdots }\otimes \mathcal{G}%
_{m}}^{(sa)}$, its trace class covering $W_{cov}$ exists and relation%
\begin{equation}
\left\Vert W\right\Vert _{cov}:=\inf_{W_{cov}\in \mathcal{T}_{\mathcal{G}%
_{1}\otimes \mathcal{\cdots }\otimes \mathcal{G}_{m}}}\mathrm{tr}[W_{cov}]
\label{02}
\end{equation}%
defines on space $\mathcal{T}_{\mathcal{G}_{1}\otimes \mathcal{\cdots }%
\otimes \mathcal{G}_{m}}^{(sa)}$ a norm, the covering norm, with properties:%
\begin{eqnarray}
\left\vert \mathrm{tr}[W]\right\vert &\leq &\left\Vert W\right\Vert
_{cov}\leq \left\Vert W\right\Vert _{1},  \label{03} \\
W\overset{\otimes }{\geq }0\text{\ \ \ } &\Rightarrow &\text{ \ \ }%
\left\Vert W\right\Vert _{cov}=\mathrm{tr}[W].  \notag
\end{eqnarray}
\end{proposition}

\section{LqHV simulation of a quantum correlation scenario}

For a state $\rho $ on $\mathcal{H}_{1}\otimes \cdots \otimes \mathcal{H}%
_{N},$ consider an $N$-partite correlation scenario\footnote{%
On the general framework for the probabilistic description of multipartite
correlation scenarios, see \cite{8}.} $\mathcal{E}_{\rho }$ where each $n$%
-th of $N$ parties performs $S_{n}$ measurements with outcomes\footnote{%
For simplicity, we consider here only discrete outcomes. This does not,
however, imply any restriction on our main results since, as it has been
proved in \cite{10}, the latter hold for outcomes of any spectral type,
discrete or continuous.} $\lambda _{n}\in \Lambda _{n}:=\{\lambda
_{n}^{(1)},...,\lambda _{n}^{(L_{n})}\}.$

We label each measurement at $n$-th site by a positive integer $%
s_{n}=1,...,S_{n}$, and each of $N$-partite joint measurements, induced by
this correlation scenario - by an $N$-tuple $(s_{1},...,s_{N})$ where $n$-th
component refers to a marginal measurement at $n$-th site.

Let, under the correlation scenario $\mathcal{E}_{\rho }$, each quantum
measurement $s_{n}$ at $n$-th site be represented on $\mathcal{H}_{n}$ by a
POV measure $\mathrm{M}_{n}^{(s_{n})}.$ For a joint measurement $%
(s_{1},...,s_{N})$ under scenario $\mathcal{E}_{\rho }$, expression 
\begin{eqnarray}
&&P_{(s_{1},...,s_{N})}^{(\mathcal{E}_{\rho })}(\lambda _{1},...,\lambda
_{N})  \label{04} \\
&=&\mathrm{tr}[\rho \{\mathrm{M}_{1}^{(s_{1})}(\lambda _{1})\otimes \cdots
\otimes \mathrm{M}_{N}^{(s_{_{N}})}(\lambda _{N})\}]  \notag
\end{eqnarray}%
specifies the joint probability $P_{(s_{1},...,s_{N})}^{(\mathcal{E}_{\rho
})}(\lambda _{1},...,\lambda _{N})$ that each $n$-th party observes an
outcome $\lambda _{n}\in \Lambda _{n}.$

If $T_{S_{1}\times \cdots \times S_{N}}^{(\rho )}$ is an $S_{1}\times \cdots
\times S_{N}$- setting source operator\footnote{%
See in section 2.} for state $\rho $, then, due to property $\mathrm{M}%
_{n}^{(s_{n})}(\Lambda _{n})=\mathbb{I}_{\mathcal{H}_{n}},$ each probability
(\ref{04}) constitutes the corresponding marginal of the normalized
real-valued distribution%
\begin{eqnarray}
&&\mathrm{tr}[T_{S_{1}\times \cdots \times S_{N}}^{(\rho )}\{\mathrm{M}%
_{1}^{(1)}(\lambda _{1}^{(1)})\otimes \cdots \otimes \mathrm{M}%
_{1}^{(S_{1})}(\lambda _{1}^{(S_{1})})\otimes  \label{05} \\
&&\cdots \otimes \mathrm{M}_{N}^{(1)}(\lambda _{N}^{(1)})\otimes \cdots
\otimes \mathrm{M}_{N}^{(S_{N})}(\lambda _{N}^{(S_{N})})\}],  \notag \\
\lambda _{n}^{(s_{n})} &\in &\Lambda _{n},\text{ \ \ }s_{n}=1,...,S_{n},%
\text{ }n=1,...,N.  \notag
\end{eqnarray}%
This implies.

\begin{theorem}
\cite{10} For every $N$-partite quantum state $\rho $ and any positive
integers $S_{1},...,S_{N}\geq 1$, each $S_{1}\times \cdots \times S_{N}$%
\emph{\ - }setting correlation scenario $\mathcal{E}_{\rho }$ admits a local
quasi hidden variable (LqHV) model 
\begin{eqnarray}
P_{(s_{1},...,s_{N})}^{(\mathcal{E}_{\rho })}(\lambda _{1},...,\lambda _{N})
&=&\dint\limits_{\Omega }P_{1}^{(s_{1})}(\lambda _{1}\mid \omega )\cdot
\ldots \cdot P_{N}^{(s_{N})}(\lambda _{N}\mid \omega )\text{ }\nu _{\mathcal{%
E}_{\rho }}(\mathrm{d}\omega ),  \label{1} \\
\text{\ }s_{1} &=&1,...,S_{1},...,s_{N}=1,...,S_{N},  \notag
\end{eqnarray}%
where $\nu _{\mathcal{E}_{\rho }}$ is a normalized bounded real-valued%
\footnote{%
Recall that, in an LHV model, measure $\nu _{\mathcal{E}_{\rho }}$ must be
positive.} measure of some variables $\omega \in \Omega $ and $%
P_{n}^{(s_{n})}(\cdot \mid \omega ),$ $\forall s_{n},\forall n,$ are
conditional probabilities.
\end{theorem}

Thus, an arbitrary $N$-partite state $\rho $ does not need to admit an $%
S_{1}\times \cdots \times S_{N}$\emph{-}setting LHV description \cite{8} \
but it necessarily admits an $S_{1}\times \cdots \times S_{N}$\emph{-}%
setting LqHV description.

\section{Bell-type inequalities}

For a general $S_{1}\times ...\times S_{N}$-setting correlation scenario $%
\mathcal{E},$ consider a linear combination%
\begin{equation}
\dsum\limits_{s_{1},...,s_{_{N}}}\left\langle \text{ }\psi _{(s_{1},\ldots
,s_{_{N}})}(\lambda _{1},...,\lambda _{N})\right\rangle _{\mathcal{E}}
\label{2}
\end{equation}%
of averages%
\begin{eqnarray}
&&\left\langle \text{ }\psi _{(s_{1},\ldots ,s_{_{N}})}(\lambda
_{1},...,\lambda _{N})\right\rangle _{\mathcal{E}}  \label{2b} \\
&:&=\dsum\limits_{\lambda _{1}\in \Lambda _{1},...,\lambda _{N}\in \Lambda
_{N}}\psi _{(s_{1},\ldots ,s_{_{N}})}(\lambda _{1},...,\lambda _{N})\text{ }%
P_{(s_{1},...,s_{N})}^{(\mathcal{E})}(\lambda _{1},...,\lambda _{N}),  \notag
\end{eqnarray}%
specified by a family $\{\psi _{(s_{1},\ldots ,s_{_{N}})}\}$ of bounded
real-valued functions on set $\Lambda :=\Lambda _{1}\times \cdots \times
\Lambda _{N}.$ For a particular choice of functions $\{\psi _{(s_{1},\ldots
,s_{_{N}})}\}$, averages in (\ref{2b}) may reduce either to joint
probabilities or to correlation functions.

In an \emph{LHV} case, any linear combination (\ref{2}) of averages
satisfies the following tight\footnote{%
A tight LHV constraint is not necessarily extreme, see \cite{1} for details.}
LHV constraints \cite{1}:

\begin{equation}
\mathcal{B}_{\{\psi _{(s_{1},\ldots ,s_{_{N}})}\}}^{\inf }\leq
\dsum\limits_{s_{1},...,s_{_{N}}}\left\langle \text{ }\psi _{(s_{1},\ldots
,s_{_{N}})}(\lambda _{1},...,\lambda _{N})\text{ }\right\rangle _{\mathcal{E}%
_{lhv}}\leq \mathcal{B}_{\{\psi _{(s_{1},\ldots ,s_{_{N}})}\}}^{\sup },
\label{3}
\end{equation}%
with the LHV constants%
\begin{eqnarray}
\mathcal{B}_{\{\psi _{(s_{1},\ldots ,s_{_{N}})}\}}^{\sup } &=&\sup_{\lambda
_{n}^{(s_{n})}\in \Lambda _{n},\text{ }\forall s_{n},\forall n}\text{ }%
\dsum\limits_{s_{1},...,s_{_{N}}}\psi _{(s_{1},\ldots ,s_{_{N}})}(\lambda
_{1}^{(s_{1})},...,\lambda _{N}^{(s_{N})}),  \label{4} \\
&&  \notag \\
\mathcal{B}_{\{\psi _{(s_{1},\ldots ,s_{_{N}})}\}}^{\inf } &=&\inf_{\lambda
_{n}^{(s_{n})}\in \Lambda _{n},\text{ }\forall s_{n},\text{ }\forall n}\text{
}\dsum\limits_{s_{1},...,s_{_{N}}}\psi _{(s_{1},\ldots ,s_{_{N}})}(\lambda
_{1}^{(s_{1})},...,\lambda _{N}^{(s_{N})}).  \notag
\end{eqnarray}

The general LHV constraint form (\ref{3}) incorporates as particular cases
both - the LHV constraints on correlation functions and the LHV constraints
on joint probabilities.

\emph{A Bell-type inequality is any of the tight linear LHV constraints (\ref%
{3}) that may be violated in a non-LHV case.}

\section{Quantum violations}

For an arbitrary $S_{1}\times \cdots \times S_{N}$-setting quantum scenario $%
\mathcal{E}_{\rho }$ specified by joint probabilities (\ref{04}), every
linear combination (\ref{2}) of its averages satisfies the following analogs 
\cite{10} of the LHV constraints (\ref{3}):%
\begin{eqnarray}
&&\mathcal{B}_{\{\psi _{(s_{1},\ldots ,s_{_{N}})}\}}^{\inf }-\frac{\Upsilon
_{S_{1}\times \cdots \times S_{N}}^{(\rho ,\Lambda )}-1}{2}(\mathcal{B}%
_{\{\psi _{(s_{1},\ldots ,s_{_{N}})}\}}^{\sup }-\mathcal{B}_{\{\psi
_{(s_{1},\ldots ,s_{_{N}})}\}}^{\inf })  \label{5} \\
&\leq &\dsum\limits_{s_{1},...,s_{N}}\left\langle \psi _{(s_{1},\ldots
,s_{_{N}})}(\lambda _{1},...,\lambda _{N})\text{ }\right\rangle _{\mathcal{E}%
_{\rho }}  \notag \\
&\leq &\mathcal{B}_{\{\psi _{(s_{1},\ldots ,s_{_{N}})}\}}^{\sup }+\frac{%
\Upsilon _{S_{1}\times \cdots \times S_{N}}^{(\rho ,\Lambda )}-1}{2}(%
\mathcal{B}_{\{\psi _{(s_{1},\ldots ,s_{_{N}})}\}}^{\sup }-\mathcal{B}%
_{\{\psi _{(s_{1},\ldots ,s_{_{N}})}\}}^{\inf }),  \notag
\end{eqnarray}%
where 
\begin{equation}
\Upsilon _{S_{1}\times \cdots \times S_{N}}^{(\rho ,\Lambda )}=\sup_{\{\psi
_{(s_{1},\ldots ,s_{_{N}})}\}}\frac{1}{\mathcal{B}_{\{\psi _{(s_{1},\ldots
,s_{_{N}})}\}}}|\dsum\limits_{s_{1},...,s_{N}}\left\langle \psi
_{(s_{1},\ldots ,s_{_{N}})}(\lambda _{1},...,\lambda _{N})\text{ }%
\right\rangle _{\mathcal{E}_{\rho }}|,  \label{5'}
\end{equation}%
is \emph{the maximal violation} by state $\rho $ of any Bell-type inequality
(either on correlation functions or on joint probabilities) specified for
settings up to setting $S_{1}\times \cdots \times S_{N}$ and outcomes in set 
$\Lambda =\Lambda _{1}\times \cdots \times \Lambda _{N}$. In (\ref{5'}), 
\begin{equation}
\mathcal{B}_{\{\psi _{(s_{1},\ldots ,s_{_{N}})}\}}:=\max \{|\mathcal{B}%
_{\{\psi _{(s_{1},\ldots ,s_{_{N}})}\}}^{\sup }|,|\mathcal{B}_{\{\psi
_{(s_{1},\ldots ,s_{_{N}})}\}}^{\inf }|\}.
\end{equation}

For short, we further refer to parameter $\Upsilon _{S_{1}\times \cdots
\times S_{N}}^{(\rho ,\Lambda )}$ as the maximal $S_{1}\times \cdots \times
S_{N}$- setting Bell violation for state $\rho $ and outcomes in $\Lambda .$

Using the new Hilbert space notions specified in section 2, we have the
following general statements.

\begin{theorem}
\cite{10} For an arbitrary $N$-partite quantum state $\rho ,$ possibly
infinite-dimensional, and any positive integers $S_{1},...,S_{N}\geq 1$, the
maximal \ $S_{1}\times \cdots \times S_{N}$- setting Bell violation $%
\Upsilon _{S_{1}\times \cdots \times S_{N}}^{(\rho ,\Lambda )}$ satisfies
relation%
\begin{equation}
1\leq \Upsilon _{S_{1}\times \cdots \times S_{N}}^{(\rho ,\Lambda )}\leq
\inf_{T_{S_{1}\times \cdots \times \underset{\overset{\uparrow }{n}}{1}%
\times \cdots \times S_{N}}^{(\rho )},\text{ }\forall n}||T_{S_{1}\times
\cdots \times \underset{\overset{\uparrow }{n}}{1}\times \cdots \times
S_{N}}^{(\rho )}||_{cov},  \label{6}
\end{equation}%
for any outcome set $\Lambda =\Lambda _{1}\times \cdots \times \Lambda _{N}.$
Here, $\left\Vert \cdot \right\Vert _{cov}$ is the covering norm and infimum
is taken over all source operators $T_{S_{1}\times \cdots \times \underset{%
\overset{\uparrow }{n}}{1}\times \cdots \times S_{N}}^{(\rho )}$ for all $%
n=1,...,N.$
\end{theorem}

\begin{corollary}
\cite{10} If a state $\rho $ has a tensor positive source operator $%
T_{S_{1}\times \cdots \times \underset{\overset{\uparrow }{n}}{1}\times
\cdots \times S_{N}}^{(\rho )}$ then it admits an $S_{1}\times \cdots \times
S_{N}$- setting LHV description for any finite number $S_{n}$ of measurement
settings at site "$n$".
\end{corollary}

\begin{corollary}
\cite{10} If a state $\rho $ has a tensor positive source operator $%
T_{S_{1}\times \cdots \times S_{N}}^{(\rho )}$, then this state admits an \ $%
S_{1}\times \cdots \times \widetilde{S}_{n}\times \cdots \times S_{N}$-
setting LHV description for any finite number $\widetilde{S}_{n}$ of
settings at each $n$-th site.
\end{corollary}

\section{Numerical bounds}

The general analytical upper bound (\ref{6}) allows us to find \cite{10} the
following new exact numerical bounds on the maximal quantum Bell violations.

\begin{itemize}
\item For the two-qubit singlet $\rho _{\text{singlet}}$, the maximal Bell
violation 
\begin{equation}
\Upsilon _{S\times 2}^{(\rho _{\text{singlet}},\Lambda )}\leq \sqrt{3},\text{
\ \ \ }S\geq 2,  \label{7}
\end{equation}%
for any outcome set $\Lambda =\Lambda _{1}\times \Lambda _{2}$, in
particular, for any number of outcomes at each site. Note that, due to the
seminal results of Tsirelson\footnote{%
Tsirelson B.: \emph{J. Soviet Math}. \textbf{36}, 557 (1987).} and Fine 
\footnote{%
Fine A.: \emph{Phys. Rev. Lett.\ }\textbf{48}, 291 (1982)}, the maximal Bell
violation $\Upsilon _{2\times 2}^{(\rho ,\Lambda )}\leq \sqrt{2}$, for any
bipartite state $\rho $ and any outcome set $\Lambda =\{\lambda _{1}^{(1)},$ 
$\lambda _{1}^{(2)}\}\times \{\lambda _{2}^{(1)},$ $\lambda _{2}^{(2)}\}$
(dichotomic measurements). The maximal violation by the singlet of any \emph{%
correlation} Bell-type inequality is given \cite{3} by the Grothendieck
constant $\sqrt{2}\leq K_{G}(3)\leq 1.5163...$ .

\item For the $N$-qudit GHZ state 
\begin{equation}
\frac{1}{\sqrt{d}}\sum_{j=1}^{d}\underset{N}{\underbrace{|j\rangle \otimes
\cdots \otimes |j\rangle }},  \label{8}
\end{equation}%
violation of any Bell-type inequality for $S$ settings and any number of
outcomes per site cannot exceed%
\begin{eqnarray}
&&\min \{(2S-1)^{N-1},\text{ }1+2^{N-1}(d-1)\}  \label{9} \\
&\leq &1+2^{N-1}\left[ \min \{S^{N-1},d\}-1\right] .  \notag
\end{eqnarray}

\item For the generalized $N$-qubit GHZ state 
\begin{equation}
\sin \varphi \text{ }|1\rangle ^{\otimes N}+\cos \varphi \text{ }|2\rangle
^{\otimes N},  \label{10}
\end{equation}%
violation of any Bell-type inequality for $S$ settings and any number of
outcomes per site is upper bounded by%
\begin{equation}
1+2^{N-1}\left\vert \sin 2\varphi \right\vert .  \label{11}
\end{equation}

\item For an arbitrary state $\rho $ on $\mathbb{C}^{d_{1}}\otimes \cdots
\otimes \mathbb{C}^{d_{N}},$ the maximal Bell violation in case of $S_{n}$
settings and any number of outcomes at each $n$-th site is upper bounded by%
\begin{equation}
1+2^{N-1}\left[ \min \left\{ \frac{S_{1}\cdot \ldots \cdot S_{N}}{%
\max_{n}S_{n}},\frac{d_{1}\cdot \ldots \cdot d_{N}}{\max_{n}d_{n}}\right\} -1%
\right] .  \label{12}
\end{equation}%
If $S_{1}=\ldots =S_{N}=S,$ then the maximal Bell violation cannot exceed%
\begin{eqnarray}
&&\min \{(2S-1)^{N-1},\text{ }1+2^{N-1}(\frac{d_{1}\cdot \ldots \cdot d_{N}}{%
\max_{n}d_{n}}-1)\}  \label{12'} \\
&\leq &1+2^{N-1}\left[ \min \{S^{N-1},\frac{d_{1}\cdot \ldots \cdot d_{N}}{%
\max_{n}d_{n}}\}-1\right] .  \notag
\end{eqnarray}%
From this $N$-partite bound it follows that violation by an arbitrary $N$%
-partite quantum state, possibly infinite-dimensional, of any Bell
inequality for $S$ measurement settings and any number of outcomes per site
cannot exceed $(2S-1)^{N-1}.$
\end{itemize}

\subsection{Bipartite and tripartite bounds}

For $N=2,$ the general upper bound (\ref{12}) implies the following \emph{%
bipartite upper bound} \cite{10} on the maximal Bell violation 
\begin{equation}
\Upsilon _{S_{1}\times S_{2}}^{(\rho ,\Lambda )}\leq 2\min
\{S_{1},S_{2},d_{1},d_{2}\}-1  \label{13}
\end{equation}%
for any quantum state $\rho $ on $\mathbb{C}^{d_{1}}\otimes \mathbb{C}%
^{d_{2}}$ and any outcome set $\Lambda =\Lambda _{1}\times \Lambda _{2}.$
This new bipartite upper bound improves:

\begin{itemize}
\item for (i) $d_{1}=d_{2}=2,$ $L_{1}=L_{2}=2,$ and (ii) $d_{1}=d_{2}\leq $ $%
L_{1}L_{2}$ $(K_{G}+1)$, $\forall L_{1},L_{2},$ the corresponding numerical
upper bounds on the maximal Bell violation (in our notation): 
\begin{eqnarray}
\text{(i) \ \ }\Upsilon _{S_{1}\times S_{2}}^{(\rho ,\Lambda )} &\leq
&2K_{G}+1,\text{ \ \ \ }L_{1}=L_{2}=2,\text{ }  \label{15} \\
\text{(ii)\ \ \ }\Upsilon _{S_{1}\times S_{2}}^{(\rho ,\Lambda )} &\leq
&2L_{1}L_{2}(K_{G}+1)-1,\text{ \ \ \ }\forall L_{1},L_{2},  \notag
\end{eqnarray}%
found in \cite{6} for any bipartite quantum state $\rho $ and $L_{1},L_{2}$
outcomes at Alice's and Bob's sites. Here, $K_{G}=\lim_{n\rightarrow \infty
}K_{G}(n)\in $ $[1.676...,1.782...]$ is the Grothendieck constant;

\item the approximate bipartite estimate 
\begin{equation}
\Upsilon _{S\times S}^{(\rho ,\Lambda )}\preceq \min \{S,d\},\text{ \ \ \ }%
\forall \Lambda ,  \label{15'}
\end{equation}%
derived in \cite{7} up to an unknown universal constant for any bipartite
state $\rho $ on $\mathbb{C}^{d}\otimes \mathbb{C}^{d}$;
\end{itemize}

For $N=3$, the general upper bound (\ref{12'}) implies the following \emph{%
tripartite upper bound} \cite{10} on the maximal Bell violation: 
\begin{equation}
\Upsilon _{S\times S\times S}^{(\rho ,\Lambda )}\leq \min \{(2S-1)^{2},\text{
}4\frac{d_{1}d_{2}d_{3}}{\max_{n}d_{n}}-3\},  \label{15''}
\end{equation}%
for any tripartite state $\rho $ on $\mathbb{C}^{d_{1}}\otimes \mathbb{C}%
^{d_{2}}\otimes \mathbb{C}^{d_{3}}$ and any outcome set $\Lambda =\Lambda
_{1}\times \Lambda _{2}\times \Lambda _{3}.$

For a state $\rho $ on $\mathbb{C}^{d}\otimes \mathbb{C}^{d}\otimes \mathbb{C%
}^{d}$, bound (\ref{15''}) implies%
\begin{eqnarray}
\Upsilon _{S\times S\times S}^{(\rho ,\Lambda )} &\leq &\min \{(2S-1)^{2},%
\text{ }4d^{2}-3\} \\
&\leq &4(\min \{S,d\})^{2}-3.  \notag
\end{eqnarray}

From (\ref{15''}) it follows -- the approximate lower estimate $\succeq 
\sqrt{d}$ found in \cite{5} for violation of some correlation Bell-type
inequality by some tripartite state on \ $\mathbb{C}^{d}\otimes \mathbb{C}%
^{D}\otimes \mathbb{C}^{D}$ is meaningful if only in this correlation
Bell-type inequality a number of settings per site satisfies relation

\begin{equation}
(2S-1)^{2}\succeq \sqrt{d}.  \label{16}
\end{equation}

\section{Conclusions}

Via some new Hilbert space notions and a new simulation approach, \emph{the
LqHV approach,} to the description of any quantum correlation scenario, we
have derived the analytical upper bound (\ref{6}) on the maximal Bell
violation by an $N$-partite quantum state. This has allowed us:

\begin{itemize}
\item to single out $N$-partite quantum states admitting an $S_{1}\times
\cdots \times S_{N}$-setting LHV description;

\item to find the new numerical upper bounds on Bell violations for some
concrete $N$-partite states generally used in quantum information processing;

\item to prove that violation by an arbitrary $N$-partite quantum state,
possibly infinite-dimensional, of any Bell inequality (either on correlation
functions or on joint probabilities) for $S$ measurement settings and any
number of outcomes per site cannot exceed $(2S-1)^{N-1}$;

\item to improve the bipartite upper bounds reported in \cite{6, 7};

\item to show that, for an "arbitrarily large" tripartite quantum Bell
violation argued in \cite{5} to be reached, not only a Hilbert space
dimension $d$ but also a number $S$ of settings per site in the
corresponding tripartite Bell-type inequality must be large and the required
growth of $S$ with respect to $d$ is given by (\ref{16}).\bigskip
\end{itemize}

\end{document}